\documentclass{ceab}   
\usepackage{epsfig}     
\usepackage{graphicx}   
\usepackage{multirow}
\usepackage{ceabbib}     
\usepackage[T1]{fontenc}
\def\gore{Oscillations of prominences}
\begin{document}
\title{Oscillations of prominences observed by MSDP and HSFA telescopes}

\author{M. ZAPI\'OR$^{1}$ and P. KOTR\v C$^{2}$
\vspace{2mm}\\
\it $^1$Astronomical Institute, University of Wroc\l aw\\
\it Kopernika 11, 51-622 Wroc\l aw, Poland\\
\it $^2$Astronomical Institute of the Academy of Science of the Czech Republic\\
\it Fri\v cova 298, 251 65 Ond\v rejov, Czech Republic
}

\maketitle

\begin{abstract}
Results of an oscillation analysis of two solar quiescent
prominences (QPs) observed with two quite different telescopes
and spectrographs are presented. For the QP of 23 September 2009
we used data from the Multi-channel Subtractive Double Pass
Spectrograph (MSDP) installed in Bia\l k\' ow Observatory which
provides H$\alpha$ two dimensional spectra with a higher spatial
resolution but with a smaller spectral resolution. The other QP
was observed on 20 August 2010 with the Ond\v rejov Multichannel
Spectrograph (HSFA-2) in 5 spectral lines and the slit-jaw imaging
system simultaneously.  Doppler velocity was analyzed in the both
the QPs. In addition, using MSDP possibilities, for the QP of 23
September 2009 we analyzed also movements in the plane of sky. In
both the QPs we found oscillations with a small amplitude (less
than 0.3 km/s) and with a medium period (approx. a dozen of
minutes) which occurred simultaneously. We discuss results
reported by other authors.
\end{abstract}


\section{Introduction}
Dense and cold plasma of solar prominences is embedded hot
corona and supported by forces of electromagnetic field and
pressure against the downward force of gravity. As was shown by
recent HINODE observations prominences have a complicated internal
structure of many thin threads where the plasma flows along
magnetic field lines in all directions. Existence of waves and
oscillations of magnetic fields must influence material that is
absorbing and emitting radiation in spectral lines. Therefore, a
study of prominence spectra obtained with a high temporal,
spectral and spatial resolution is an important tool for
diagnostics of processes both in the prominence magnetoplasma.

The aim of this paper is to show possibilities of the two quite
different spectral and imaging solar telescopes and specrographs
for long-time observations of prominences and study of
oscillations detectable in Doppler velocity component and in the
plane of sky.

\section{Observations and Data Reductions}

The analyzed prominence data were obtained at the Multi-channel
Subtractive Double Pass Spectrograph (MSDP) installed in Bia\l k\'
ow Observatory and at the Multichannel Spectrograph (HSFA-2) in
the Ond\v rejov observatory in frame of a bilateral cooperation.
A quite numerous set of observations has been done but only a
small fraction was processed until now. For processing we tried to
select QPs both  belonging to the polar crown and those from
lower latitudes nearby active regions. Neither the orientation of
the prominence axis to the line-of-sight, nor the prominence type, 
were taken into account in the event selection.

\subsection{MSDP}
\begin{figure}[htbp]
  \begin{center}
   \epsfig{file=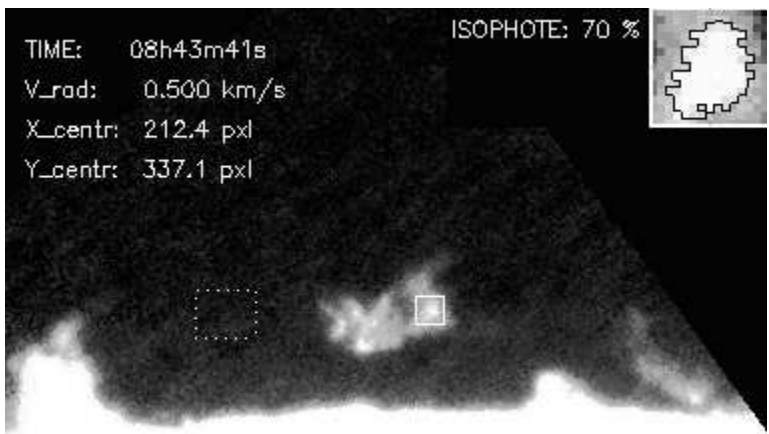,width=10cm}%
  \end{center}
  \caption{A sample image of the prominence observed on 23 September 2009 by MSDP on the eastern limb of the Sun near the active region NOAA 11026. A magnified part of the analyzed kernel is shown in the upper-right corner. White line represents the isophote which delimits analyzed area of the kernel. Dotted rectangle shows the place where the scattered light was sampled. Time of observation, calculated mean Doppler velocity and coordinates of the kernel are given in the upper-left corner.}
\label{msdp_prom}
\end{figure}

\begin{figure}[htbp]
  \begin{center}
   \epsfig{file=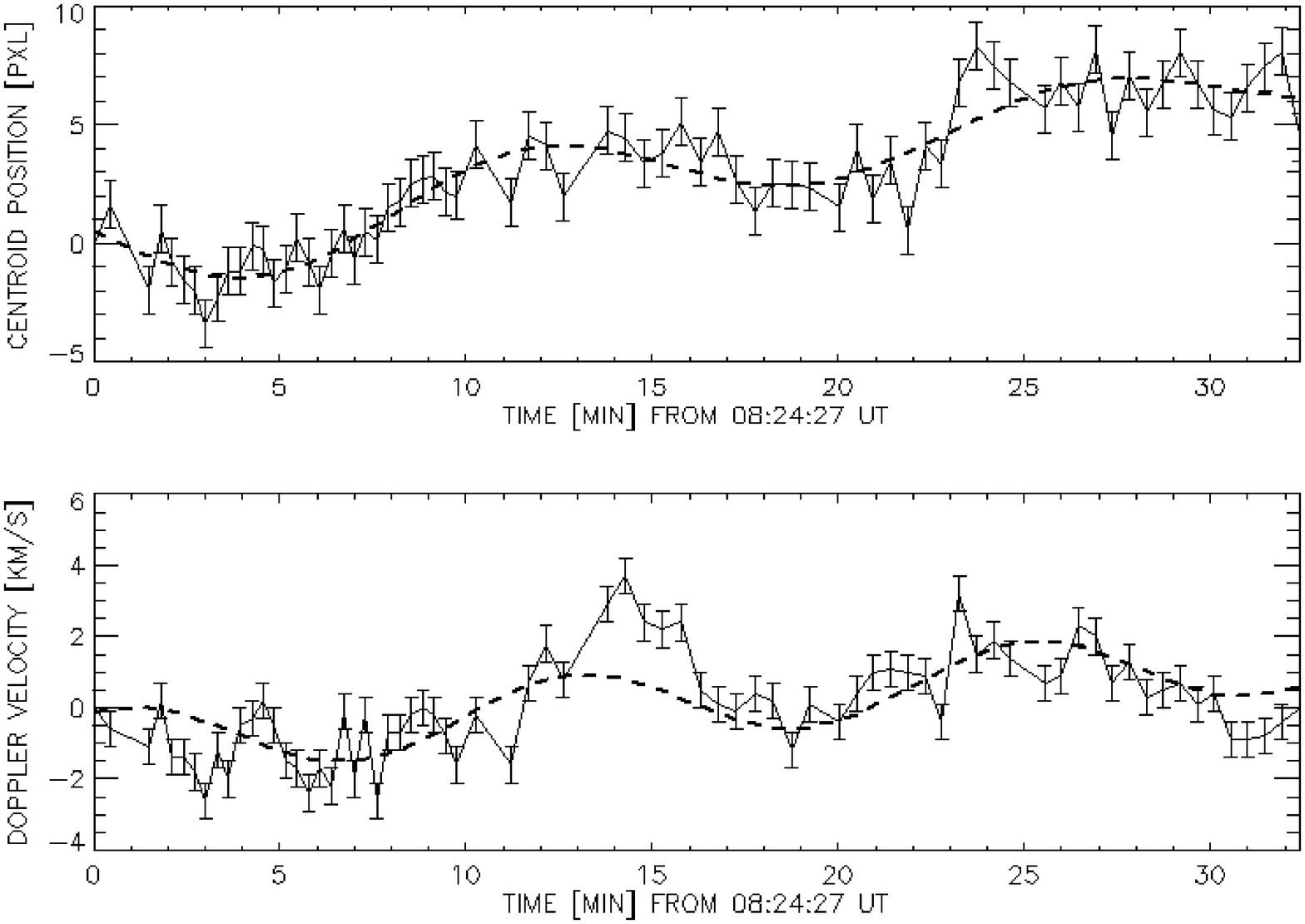,width=11cm}%
  \end{center}
  \caption{Top panel: Time sequence of position of the centroid of the analyzed kernel. Vertical axis is scaled in pixels. Bottom panel: Doppler velocity as measured from the MSDP observations. Vertical axis scaled in km/s. In both panels horizontal axis is scaled in minutes after the beginning of observations. Solid line: observed raw data points with error bars which were arbitrary set to 1 pixel in centroid position and 0.5 km/s in Doppler velocity \citep{rudawy}, dashed line: best fit with function $A\sin(2\Pi t / P + \phi)+Bt +C$, where: $A$ - amplitude, $P$ - period (estimated from Scargle periodogram), $t$ - time, $\phi$ - phase, $B,C$ - coefficients of linear trend. }
\label{xv_graph}
\end{figure}

The prominence of 23 September 2009 was observed with the
Multichannel Subtractive Double Pass Spectrograph (MSDP) installed
in Bia\l k\' ow Observatory \citep[for instrument
description and data reduction overview, see.][]{zapior07, zapior10} during the interval
08:27:24 - 09:00:00 UT. 
It was located at the south-east limb at heliographic latitude S50, about 40 degrees east of the active region NOAA 11026. 
The prominence axis was oriented almost perpendicular to the line-of-sight.
An image of the prominence observed by MSDP
at 08:43:41 UT is shown in Figure~\ref{msdp_prom}. Time cadence was
between 25 and 34 seconds depending on the size of scanning area.
After a standard data reduction (dark current subtraction,
flat-fielding, geometry correction) we chose for an analysis a
kernel in the upper-right part of the prominence (see Figure~\ref{msdp_prom}).
Due to a large amount of data we applied an automating kernel
tracking \citep[described in][]{zapior10} to point position of the
kernel in all images. Using possibilities of the MSDP we were able to obtain the mean Doppler velocity of the kernel during time of observations together with position of the centroid of the kernel in all (74) images in which the kernel was visible. During the data reduction we applied a procedure of scattered light subtraction. We took a sample of scattered light from an area in the image above the limb without any emission structure. As the sample was taken from the same image (with the same observational conditions and exposure time) we applied subtraction of mean scattered-light profile from the mean
 emission profile of the kernel directly. Subtraction of scattered light allows to increase relative peak in emission profile of the prominence. Then the emission profile was fitted by a gaussian. From shift of the center of gaussian we obtained mean a Doppler velocity of the kernel in all images. Centroid of the kernel was calculated as a centroid of all points inside the limiting isophote weighted by a pixel signal. Results are presented in Figure \ref{xv_graph}.

\subsection{HSFA-2}
One of the two horizontal solar telescopes with spectrograph HSFA2
(Ho\-ri\-zontal Sonnen Forschung Anlage) was
modified recently \citep[for telescope description, see][]{kotrc10}
for Multichannel spectrograph. It provides simultaneous
observations in five spectral lines (H$\alpha$, H$\beta$, D3 of
HeI, CaII 8542~\AA\ and CaII H or CaII K) and H$\alpha$ slit-jaw
images. Data are recorded by fast CCD cameras of 1024 x 1280 image
size with a quadratic pixel of 6.7 $\mu$m and binned to 1024 x
1280 images. Data reduction contains a standard procedure
(dark-frame and flat-fielding) as well as determination of
dispersion curve for all spectral images. The scattered light
subtraction after the calibration process was also applied to all
spectral images in all the spectral lines mentioned above.

\begin{figure}
  \begin{center}
   \epsfig{file=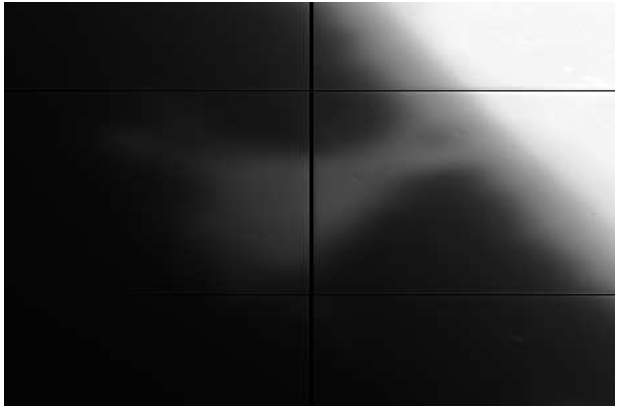,width=8cm, bb = 14 14 477 322}
  \end{center}
  \caption{Slit-jaw image of the prominence of 20 August 2010 observed by HSFA-2 at 08:53:42 UT. Vertical line represents the position of the entrance slit of the spectrograph. Horizontal lines represent hairs used for coalingment of spectral images with slit-jaw image taken with a narrow band H$\alpha$ filter.}
\label{hsfa_prom}
\end{figure}

The prominence of 20 August 2010 was observed with HSFA-2 during the interval 08:00:01 - 11:00:19 UT. It was located at the south-east limb at a heliographic latitude of 30S. 
The prominence axis was oriented almost perpendicular to the line-of-sight and can be characterized as a funnel type. An image of the prominence observed by HSFA at 08:53:42 UT is shown in Figure~\ref{hsfa_prom}. Time cadence was 9 seconds.

\begin{figure}[htbp]
  \begin{center}
   \epsfig{file=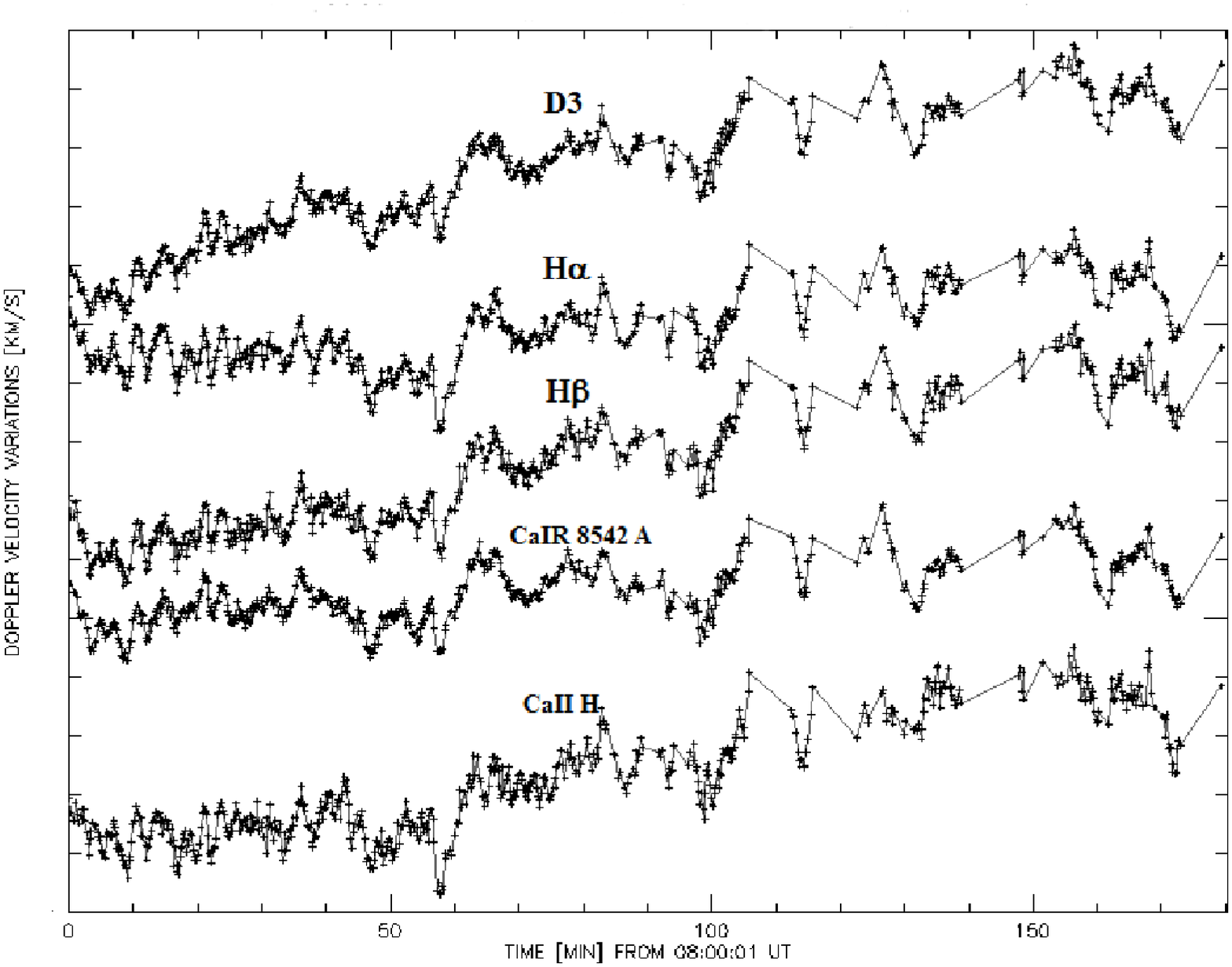,width=12cm}
  \end{center}
  \caption{Time sequence of relative Doppler velocity variations of prominence observed by HSFA-2 on 20 August 2010  during the interval 08:00:01 - 11:00:19 UT. On vertical axis one tick mark corresponds to 1 km/s, the horizontal axis is scaled in minutes after the beginning of observations. Particular measurements of individual spectral line are indicated by small crosses. }
\label{dopp}
\end{figure}

The next step was an approximation of the prominence emission
profile by a gaussian and determination of the Doppler velocity
component. The time profile of the calculated Doppler velocity (Figure \ref{dopp}) was
used for analysis and detection of oscillations.

\section{Results}
Using MSDP we detected (quasi-)periodic oscillations in Doppler
velocity component as well as in  the position of the kernel in
the plane of sky. After calculating Scargle's periodogram \citep{scargle82} from
data between 8:27 to 9:00 UT we found a period equal to 12.11
minutes in Doppler velocity and 14.35 minutes in position in
horizontal direction. No oscillations were detected in vertical direction.
Table \ref{tab:est1} summarizes obtained results.

\begin{table}
\caption{Estimation of periods found in data. $H$ - oscillations in horizontal direction, $D$ - Doppler velocity oscillations.}
\label{tab:est1}
\begin{center}
\begin{tabular}{ccc}
\hline

Spectral line & Period [min] & Amplitude \\
\hline

 \multicolumn{3}{c}{\footnotesize{\textbf{23 Sep 2009 - MSDP}}} \\
H$\alpha^D$ & 12.11 & 0.96 km/s  \\
H$\alpha^H$ & 14.35 & 1.43 pxl  \\
\hline
 \multicolumn{3}{c}{\footnotesize{\textbf{20 Aug 2010 - HSFA-2}}}\\
\multirow{3} {0.5cm}{H$\alpha^D$}    & 66.74 & 0.269 km/s    \\
  & 30.19 & 0.153 km/s    \\
   & 12.68 & 0.181 km/s    \\
  & 10.31 & 0.231 km/s    \\
   & 3.51 & 0.102 km/s    \\
\hline
\multirow{3} {0.5cm}{ Ca$^D$ }
         & 142.81 & 0.179 km/s    \\
         & 63.66 & 0.278 km/s    \\
         & 38.66 & 0.143 km/s    \\
         & 12.56 & 0.184 km/s    \\
         & 10.31 & 0.238 km/s    \\
\hline
\multirow{5} {0.5cm}{ D3$^D$  }       & 39.77 & 0.309 km/s   \\
         & 22.72 & 0.226 km/s   \\
         & 14.26 & 0.176 km/s   \\
         & 12.45 & 0.101 km/s   \\
         & 10.31 & 0.141 km/s   \\
\hline
\multirow{5} {0.5cm}{ H$\beta^D$ }
    & 66.74 & 0.177 km/s     \\
    & 39.77 & 0.242 km/s     \\
    & 21.99 & 0.178 km/s     \\
    & 12.68 & 0.148 km/s     \\
    & 10.31 & 0.186 km/s     \\
\hline
\multirow{5} {0.8cm}{ CaIR$^D$  }    & 40.95 & 0.263 km/s    \\
      & 22.35 & 0.194 km/s    \\
      & 14.56 & 0.187 km/s    \\
      & 12.56 & 0.123 km/s    \\
      & 10.31 & 0.160 km/s    \\
\hline
\end{tabular}
\end{center}
\end{table}

The HSFA-2 data were influenced by seeing. It causes shifting of 
the image of the prominence on the spectrograph slit and as a
result we observe slightly different parts of the prominence.
Fortunately, exposure time for prominence spectra was about 100 ms
and the seeing was approx. 2 - 3 arcsec. In the axis parallel to
the slit we applied coalignment of slit-jaw image as well as
spectral images using correlations of visible structures. After
that we calculated a shift in perpendicular axis. Finally we
rejected from further analysis images with shift greater than 10
pixels (2 arcsec). From fitting gaussian to emission profiles of
the prominence we calculate Doppler velocity time series which
were analyzed using Scargle's formula. We
detected peaks in periodograms of particular time series for
spectral line observed by HSFA-2 (see Table \ref{tab:est1}).
Values of detected periods slightly vary from line to line,
however periods 10.31 min and 12.45-12.68 min occur in all lines.
We conclude that only these two periods could be treated as real
since there is no reason to oscillate mass in particular
prominence with different periods in different lines. Detected
amplitudes are in the order 0.1-0.2 km/s.

\section{Discussion and Conclusions}
Doppler velocity oscillations were found in both analyzed
prominences. From MSDP data of prominence of 23 September 2009 we
found period 12.11 min and from HSFA-2 periods 10.31 min and
$\approx$12.56 min. Although the most of observed periods are
grouped in two classes: less than 10 minutes and more than 40
minutes \citep{oliver02}, periods in the range 10-40 min were 
also observed recently  \citep[e.g.,][]{yi91, sutt97, reg2001aap}. Amplitudes detected by us are rather low
($\approx$ 1.0 km/s in MSDP case and $<$0.3km/s in HSFA-2 case),
however other authors also reported such low values
\citep[e.g.,][]{mash84}. For classification of
prominence oscillations, see also \citet{vrsnak1993}. 

Taking to the account period detected in
the plane of the sky from MSDP data one can estimate magnetic
field geometry. If we assume that prominence axis lays exactly in the plane of the sky then 
according \citet{anzer2009} there is:
$$P_{SKY}/P_{Dopp}=B_z/B_x=\tan \alpha$$ where:
$P_{SKY}$, $P_{Dopp}$ - periods in the plane of the sky and in
Doppler signal, $B_z$, $B_x$ - 
components of magnetic field (vertical and parallel to prominence axis respectivately), 
$\alpha$ - inclination of magnetic field lines
on the prominence boundary. From our estimations there is:
$P_{SKY}/P_{Dopp}=14.35/12.11\approx 1.2$ which yields to $\alpha
\approx$ 50 degrees.

Future analysis of data of prominences observed simultaneously by
MSDP and HSFA-2 during observational campaign in August 2010 may
bring new results concerning prominence oscillations. Then the
numerous set of observations will allow us to apply selection 
criteria concerning types, orientations, and positions of analyzed 
quiescent prominences.

\medskip

\section*{Acknowledgements}

PK appreciates support of the GA CR grants  205/09/1469,
P209/10/1706 and the Astronomical Institute Research Project
AV0Z10030501. MZ would like to thank prof. Pawe\l ~Rudawy for his help 
during preparation of this article.

\bibliographystyle{ceab}

\bibliography{zapior_corr}

\end{document}